# ARTICLE

# Photochemical reaction on graphene surfaces controlled by substrate-surface modification with polar self-assembled monolayers†

Ryo Nouchi *[a,b] and Kei-ichiro Ikeda [a]

The unique thinness of two-dimensional materials enables control over chemical phenomena at their surfaces by means of various gating techniques. For example, gating methods based on field-effect-transistor configurations have been achieved. Here, we report a molecular gating approach that makes use of a local electric field generated by a polar self-assembled monolayer formed on the supporting substrate. By performing Raman scattering spectroscopy analyses with a proper data correction procedure, we found that molecular gating is effective for controlling solid phase photochemical reactions of graphene with benzoyl peroxide. Molecular gating offers a simple method to control chemical reactions on the surfaces of two-dimensional materials because it requires neither the fabrication of a transistor structure nor the application of an external voltage.

## 1 Introduction

Atomically thin two-dimensional (2D) materials can be simply obtained from exfoliation of layered compounds.[1-3] Such uniquely thin materials allow for tuning of the charge carrier concentration in 2D materials by means of electrostatic gating in a field effect transistor (FET) configuration. A FET is an electronic device that has the same structure as a parallel-plate capacitor; the charge carrier density in one plate (typically a semiconductor) can be tuned by applying a gate voltage to the other plate (i.e., the gate electrode), and the two plates are separated by an insulator. FET gating has been used to control various phenomena where the electron/hole concentration play an important role, such as in insulator-to-metal/superconductor transitions,[4-7] adsorption of foreign molecules,[8,9] and surface chemical reactions.[10-13]

The carrier concentration in 2D materials can be controlled through methods other than FET gating. For example, surface charge transfer from adsorbed atoms/molecules has been used to tune the carrier concentration in 2D materials.[14-18] However, tuning based on surface charge transfer requires deposition of foreign atoms/molecules onto the surfaces of 2D materials, which thus makes it impossible to use this method to retain control of surface phenomena. Such 2D materials possess very high surface-to-volume ratios because of their low thickness, thus surface-related phenomena in these materials have been widely explored.[8-13, 19-23]

Another compatible method is surface modification of the supporting substrate with a self-assembled monolayer (SAM) consisting of molecules that have a permanent electric dipole. A local electric field generated by the dipoles from the constituent molecules electrostatically dope charge carriers to a solid put on the SAM-modified substrate. The concentration of charge carriers in the solid can be controlled with the use of a SAM molecule having a different orientation/magnitude of its electric dipole moment. This method has been used to control the threshold voltage of organic field-effect transistors,[24, 25] and more recently to control the carrier concentration of 2D materials, such as graphene[26-28] and transition metal dichalcogenides.[29]

In this study, a chemical reaction on graphene surfaces has been found to be controllable by modifying the substrate surface with SAMs of polar molecules. The electric dipoles of the constituent molecules in the SAM change the charge carrier concentration in the graphene over-layer, which provides an electrostatic control of carrier concentration as the same as FET gating. In this paper, we call it molecular gating to distinguish it from FET gating. Molecular gating is tested with a solid phase photochemical reaction with benzoyl peroxide (BPO), which is one of representative molecules for modification of graphene surfaces.[12, 30, 31] This methodology is indeed found to be effective for controlling the reaction as evidenced by the doping-level-dependent degree of the photochemical reaction, which is obtained after properly following the correction procedure of Raman data.[32, 33] Molecular gating offers a simple configuration that does not require an external (gate) voltage and could be widely applied to other chemical reactions and surface phenomena.

[a.] Department of Physics and Electronics, and Nanoscience and Nanotechnology Research Center, Osaka Prefecture University, Sakai 599-8570, Japan. E-mail: r-nouchi@pe.osakafu-u.ac.jp
[b.] PRESTO, Japan Science and Technology Agency, Kawaguchi 332-0012, Japan

† Electronic Supplementary Information (ESI) available: Correction of Raman data, a spectrum of the light source used in this study, and Raman scattering spectra from the graphene samples. See DOI: 10.1039/x0xx00000x

## 2 Experimental

### A. Sample preparation

A highly doped Si wafer with a thermally grown 300-nm thick oxide layer was used as the supporting substrate for graphene. The Si substrate was cleaned with acetone and isopropanol in an ultrasonic bath, followed by an oxygen plasma treatment (Harrick Plasma, PDC-32G). The substrate was then immersed in a 2 wt% hexane solution of SAM molecules for 20 h under ambient conditions. The molecules used to form the SAMs in this study are *n*-propyltriethoxysilane (Tokyo Chemical Industry, purity > 98%) and 1*H*,1*H*,2*H*,2*H*-perfluorooctyltriethoxysilane (Tokyo Chemical Industry, purity > 95%); SAMs fabricated from these molecules are hereafter denoted $CH_3$-SAM and F-SAM, respectively. 3-Aminopropyltriethoxysilane (Nacalai Tesque, purity > 98%) was also used to check the water contact angle, and the fabricated SAM from this molecule is hereafter called $NH_2$-SAM; a 2 wt% acidic ethanol (1 mM acetic acid in ethanol) solution was used to prepare $NH_2$-SAM.[26] The substrate was removed from the solution, dried in air, and subsequently cleaned with pure hexane or ethanol in an ultrasonic bath. Water contact angles on the SAM-modified substrates were measured immediately after the SAM formation with distilled water with the use of a contact angle meter (Excimer, SImage standard 100II). Graphene flakes were formed by mechanical exfoliation with adhesive tape and deposited on the substrate surface immediately after modification with the SAM.

### B. Raman scattering spectroscopy characterization

The number of layers and progress of chemical reactions were determined using a Raman microscope (Nanophoton, Raman DM) equipped with a 532-nm green laser. Wavenumbers of the acquired Raman spectra were corrected to the Si peak at 520 cm$^{−1}$. To minimize the effect of the intra-flake inhomogeneity,[34] an areal average of the Raman spectra was taken over each flake. The area for taking the average was chosen to avoid the edges of graphene flakes, where a finite intensity of the D band was expected before the chemical reaction.

### C. Photochemical reaction

The SAM-modified substrate with graphene flakes was immersed in a 10 mM acetone solution of BPO (Nacalai Tesque, contains 25 vol% water) for 30 min. The substrate was removed from the solution and dried with air. After deposition, BPO was discernible to the naked eye as a white film on the substrate. The sample was irradiated with UV light for 10 min in air with a UV irradiation system (Ushio, SP7-250UB); the spectral irradiance is shown in the ESI, Fig. S1) equipped with a deep UV lamp (Ushio, UXM-Q256BY). The irradiance at the sample was ca. 0.8 W cm$^{−2}$. The spot size of the UV light was sufficiently large to completely irradiate the entire graphene surface. To prevent the Raman laser from inducing photochemical reactions, the unreacted BPO film was cleaned by immersion in acetone for 30 min before acquiring the Raman scattering spectra; the cleaning period of 30 min was determined by careful examinations with different cleaning periods.

## 3 Results and discussion

### A. Controlled carrier concentration

Fig. 1(a) shows the structure of the sample used in this study. The $SiO_2$/Si substrate surface was modified with a SAM of polar molecules via silane coupling. The chemical structures of the constituent molecules in the SAM are shown in Fig. 1(b) along with the direction of their electric dipole moments. In previous studies on carrier concentration control,[24, 26] SAM molecules with an amino group ($NH_2$-SAM) were used in addition to $CH_3$ and F-SAM molecules. We measured the water contact angles on the SAM-modified substrates to be 80.3°, 86.9°, and 33.0° for $CH_3$-SAM, F-SAM, and $NH_2$-SAM, respectively. The substrate surface treated with the $NH_2$-SAM became hydrophilic in contrast to the hydrophobic surfaces obtained by treatment with $CH_3$-SAM or F-SAM. In general, 2D materials are partially transparent in terms of their surface wetting properties, thus graphene deposited on a hydrophilic surface is more hydrophilic than that on a hydrophobic surface.[35] The wettability will also affect the degree of chemical modification. For instance, photo-oxidation of graphene has been shown to be markedly affected by the presence of water on the supporting substrate surface[36] and the humidity of the ambient environment.[13] Furthermore, photo-transformations of $WS_2$ are also known to be affected by ambient moisture.[37] To avoid the possible complexity introduced by differences in surface wettability and directly investigate the effects of carrier concentration, we used SAM molecules that formed hydrophobic surfaces, i.e., $CH_3$-SAM and F-SAM molecules.

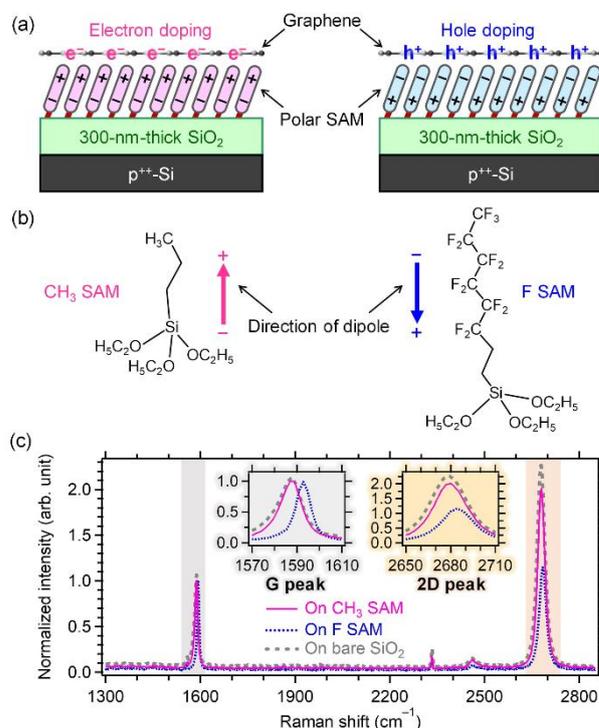

Fig. 1 Carrier concentration control in graphene by modification of the supporting substrate surface with polar SAMs. (a) Schematic diagram of the carrier concentration control by the orientation of SAM molecules. (b) Chemical structure of the SAM molecules used in this study. Arrows indicate the orientation of the permanent electric dipole. (c) Typical Raman scattering spectra of graphene formed on CH$_3$-SAM and F-SAM, showing the median G peak position among all tested flakes. For the reference, a Raman spectrum of graphene formed on a bare SiO$_2$ substrate is also shown. The spectra are normalized to the peak intensity of the G peaks. Insets show enlarged views of the G and 2D band regions.

Raman scattering spectroscopy is a powerful tool for characterizing the carrier doping level in graphene. Fig. 1(c) shows typical Raman scattering spectra of graphene deposited on SAM-modified and bare substrates. Characteristic Raman peaks of graphene were observed near 1590 cm$^{-1}$ (G band) and 2680 cm$^{-1}$ (2D band). The peak frequencies exhibit a shift that depends on the nature of the SAM molecules. One can determine the doping level of the graphene from the relationship between the frequencies of the two peaks.[32] The difference in the carrier doping level can also be measured from changes in the 2D to G peak intensity ratio,[38] which serves as further evidence that the carrier concentration can be controlled by choosing the appropriate polar SAM. The Raman spectrum of graphene formed on CH$_3$-SAM is similar to that on a bare substrate, but that on F-SAM exhibited a shift in the G-peak frequency and a reduction in the 2D-to-G ratio. Therefore, we confirmed that the carrier doping level was indeed altered by the SAM on the supporting substrate surface. Although no metal electrode was fabricated on the exfoliated flakes, single-layer graphene used in this study was found inside a flake whose thickness varies within the flake, and thick multilayer graphene acted as an electron reservoir that doped charge carriers into the single-layer portions;[26] besides, molecular adsorbates from the air (such as water and oxygen) also acted as the reservoir. These carrier-controlled flakes were then used to control a chemical reaction at the graphene surfaces.

**B. Photochemical reaction**

Fig. 2(a) shows an experimental procedure for the solid phase photochemical reaction with BPO. After deposition of a BPO film on graphene, the film was irradiated with UV light for 10 min under ambient conditions. BPO is known to photodissociate to generate a phenyl radical, which subsequently attaches to the graphene surface. Fig. 2(b) shows typical Raman scattering spectra from the BPO-covered graphene film after UV irradiation. An additional peak near 1340 cm$^{-1}$ appeared on the CH$_3$-SAM-modified substrate after UV irradiation. This peak is the D band of graphene and is known to be a measure of the progress of surface chemical reactions.[39] As shown in Fig. 2(b), a clear D band appeared only in graphene on the CH$_3$-SAM-modified substrate, which shows that the reactivity in graphene on the CH$_3$-SAM-modified substrate is higher than that on the F-SAM-modified substrate. Therefore, molecular gating can be used to control the solid-phase photochemical reaction with BPO.

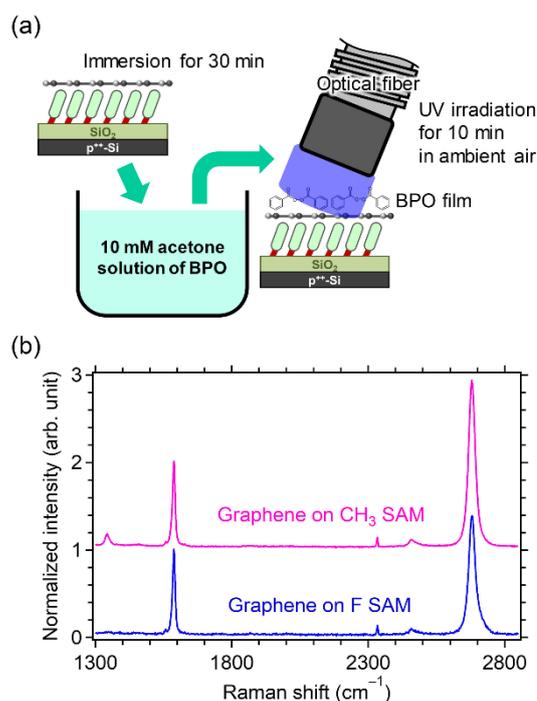

Fig. 2 SAM-controlled solid phase photochemical reaction between graphene and BPO. (a) Experimental procedure. (b) Raman scattering spectra of graphene after the reaction, which shows the largest areal intensity of the D peak among all tested flakes. Spectra are normalized to the G peak intensity. The spectrum of graphene on CH$_3$-SAM is shifted for clarity.

## 4 Discussion

**A. Data correction procedure**

The degree of the chemical reaction can be expressed with the use of the D-to-G intensity ratio of the Raman spectrum.[33, 39-42] However, the D-to-G ratio is known to be dependent on the charge carrier concentration in graphene.[33, 40, 41] Thus, a direct comparison of the as-acquired Raman data should be avoided when comparing two flakes with different doping levels. Instead, the as-acquired Raman data should be corrected by examining the carrier doping level of each flake. Although the carrier concentration can be determined by inspecting the peak position of the G band,[33] unintentional strain also shifts the G peak wavenumber.[43] To separate the effects of doping and strain on the G peak wavenumber, we inspected the correlation between the wavenumbers of the G and 2D peaks.[32] The strain- and doping-induced changes in the 2D–G correlation have a linear relationship; the slopes of these lines, i.e., $\Delta\omega_{2D}/\Delta\omega_G$, have been reported to be 2.2 for a uniaxial strain,[32] and 0.55 (0.2) for hole (electron) doping.[33] The electron doping line is known to show a deviation from the slope of 0.2 when $\Delta\omega_G$ becomes higher than ~10 cm$^{-1}$.[33] Here, $\Delta\omega_{2D}$ and $\Delta\omega_G$ are respectively defined as: $\omega_{2D} - \omega_{2D}^0$ and $\omega_G - \omega_G^0$, where $\omega_{2D}$ ($\omega_G$) is the wavenumber of the 2D (G) peak under investigation, and $\omega_{2D}^0$ ($\omega_G^0$) is that with no strain and no doping. Therefore, if $\omega_{2D}^0$ and $\omega_G^0$ are known, the $\Delta\omega_G$ value solely from the carrier doping effect can be determined by following the vector decomposition analyses[32, 33] shown in Fig. 3(a).

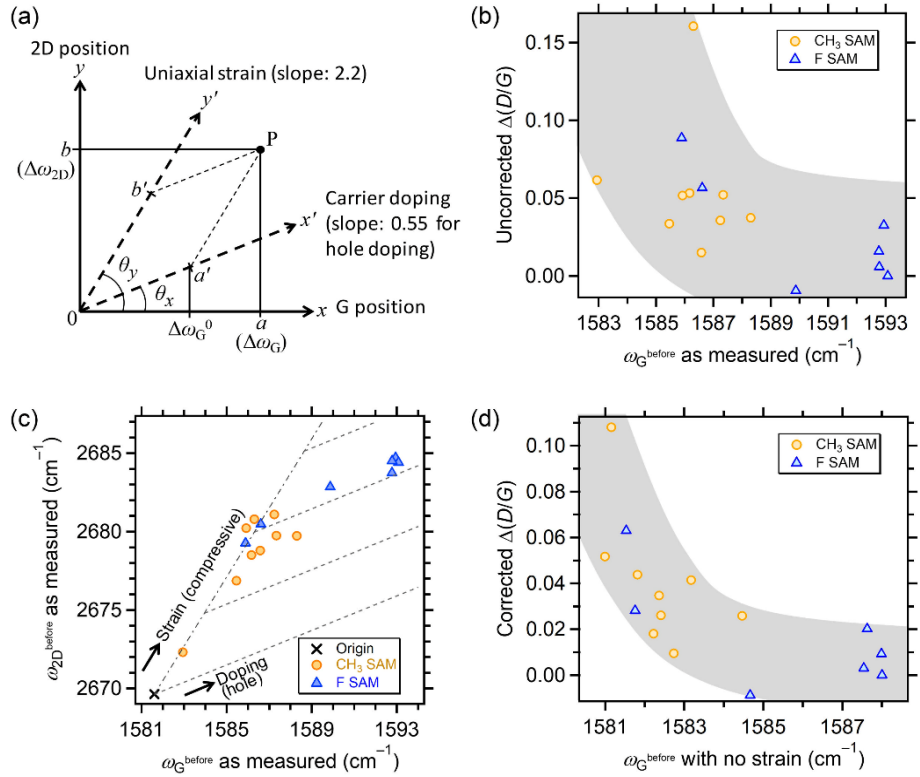

Fig. 3  Data correction for the relationship between the degree of solid phase photochemical reaction with BPO and the SAM-controlled doping level of graphene. (a) Vector decomposition analysis for extracting strain-free values of the G-peak wavenumbers. (b) Uncorrected relationship with as-measured values. (c) 2D–G correlation of the samples measured in this study. (d) Corrected relationship between the reaction degree corresponding to no doping and the G-peak wavenumbers solely determined by the doping level. Each data point in (b), (c), and (d) is from a different flake by taking areal averages of Raman spectra of each flake. Gray lines in (b) and (d) indicate the data trend, which is clearer after the data correction procedure.

From the $\Delta\omega_G$ value solely from the carrier doping effect, which was obtained by following the vector decomposition analysis[32, 33] (see the ESI for details), the Fermi level relative to the Dirac point, $E_F$, can be calculated[33] as described in the ESI. The $E_F$ value is used to determine a carrier-concentration dependence of the D-to-G ratio. According to Froehlicher and Berciaud,[33] the carrier-concentration dependence of the D-to-G ratio is similar to that of the 2D-to-G ratio, and follows the relationship:

$$\left.\frac{A_D}{A_G}\right|_0 = \frac{A_D}{A_G}\left(1 + \frac{0.06|E_F|}{\gamma_{e\text{-Ph}} + \gamma_D}\right)^2,$$

where $A_D$ and $A_G$ are the areal (integrated) intensity of the G and D bands, respectively; $A_D/A_G|_0$ corresponds to the undoped value (i.e., $E_F = 0$; at the Dirac point); $\gamma_{e\text{-Ph}}$ and $\gamma_D$ are the electron-phonon and electron-defect scattering rates, respectively. The value $\gamma_{e\text{-Ph}}$ has been reported to be 47 ± 7 meV, and $\gamma_D$ was found to be as small as 5 meV even if $A_D$ is comparable to the areal intensity of the 2D band.[33] In our samples, the D band intensity is smaller than the 2D band intensity, which ensures a small $\gamma_D$ value. Within the $A_D$ values obtained in this study ($A_D/A_G < 0.2$), it would be a good approximation to set $\gamma_{e\text{-Ph}} + \gamma_D$ as 50 meV. To exclude the contribution of the carrier-concentration dependence, the undoped value ($A_D/A_G|_0$) should be used instead of the as-measured value ($A_D/A_G$).

**B. Doping dependence of the reactivity**

We performed the same test with multiple flakes. The results are collected in Fig. 3(b) as a plot of $\Delta(A_D/A_G)$ versus the as-measured G-peak wavelength before the reaction procedure, $\omega_G^{before}$. $\Delta(A_D/A_G)$ is a measure of the degree of the reaction, and is defined as $A_D^{after}/A_G^{after} - A_D^{before}/A_G^{before}$, where $A_D^{after}$ ($A_G^{after}$) and $A_D^{before}$ ($A_G^{before}$) are the areal intensity of the D peak (G peak) after and before the reaction, respectively. Although the data trend seems to indicate that $\Delta(A_D/A_G)$ becomes larger with lower $\omega_G^{before}$, the data plots are very scattered at present.

Then, we followed the data correction procedure as explained above. Fig. 3(c) shows a plot of the as-measured 2D-peak wavelength, $\omega_{2D}^{before}$, versus the as-measured G-peak wavelength, $\omega_G^{before}$, before the reaction procedure. The origin ($\omega_G^0$, $\omega_{2D}^0$) is (1581.6, 2669.7) [cm$^{-1}$] as determined earlier. The $\omega_{2D}^{before}$ and $\omega_G^{before}$ values were extracted from areal-averaged spectra (all the spectra are shown in the ESI, Figs. S2 and S3). A complete data set showing $\omega_{2D}^{before}$ and $\omega_G^{before}$ values at every point in each flake (i.e., $\omega_{2D}^{before}$ and $\omega_G^{before}$ values before taking the areal average) are shown in the ESI, Fig. S4(a).

The vector decomposition analysis uses 2D–G correlation lines with a different slope, $\Delta\omega_{2D}/\Delta\omega_G$, for uniaxial strain, hole doping, and electron doping. Thus, before conducting the vector decomposition analysis, the carrier type (hole or electron) doped by the molecular gating should be determined. The slope is 2.2 for the uniaxial strain, which is considerably higher than the slopes of the doping lines (0.55 for hole doping, 0.2 for electron doping). Thus, a distributed range of $\omega_G^{before}$ is extended if the carrier doping effect dominates the $\omega_G^{before}$ distribution. The intra-flake $\omega_{2D}^{before}$–$\omega_G^{before}$ correlations of samples that showed a notable $\omega_G^{before}$ distribution were fitted with a straight line, which had slopes of 0.50 and 0.66 for single-layer flakes on CH$_3$- and F-SAM, respectively [see Fig. S4(b) in the ESI]. The obtained slopes were close to that for hole doping (0.55), which indicates that the flakes formed on CH$_3$- and F-SAM were both hole-doped. If we consider only the orientation of the permanent dipole of the SAM constituent molecule, the CH$_3$-SAM might be expected to dope electrons, as shown in Fig. 1(a). However, all the experiments were conducted in ambient air, and thus, molecules from the ambient environment, such as oxygen and water will adsorb to the graphene surface and are known to dope holes into graphene.[48, 49] In transfer characteristics of graphene FETs formed on methyl-group-based SAMs which are similar to CH$_3$-SAM in the present study, the gate voltage corresponding to the charge-neutrality point has been found to be positive, indicating that the graphene channels were hole-doped.[28] Therefore, it is most likely that the graphene flakes formed on the CH$_3$-SAM are slightly hole-doped.

Because all the flakes under investigation are most likely hole-doped, vector decomposition analysis was performed with a uniaxial strain line (slope: 2.2) and the hole doping line (slope: 0.55). These results are shown in Fig. 3(d). The horizontal axis shows the corrected G-peak wavenumber before the reaction, $\omega_G^{before}$, based on vector decomposition analysis, which corresponds to no-strain conditions. For correction of the $\Delta(A_D/A_G)$ values in the vertical axis, the D-to-G ratios before ($A_D^{before}/A_G^{before}$) and after ($A_D^{after}/A_G^{after}$) the reaction were both corrected by considering the carrier concentration dependence of the D-to-G ratio, where the doping level, $E_F$, was determined from the corrected G-peak wavenumbers before and after the reaction, respectively. A relationship between the corrected $\Delta(A_D/A_G)$ value, which corresponds to the undoped value, and the corrected $\omega_G^{before}$ is more clearly discernible in Fig. 3(d) than in the uncorrected plots in Fig. 3(b). Thus, the degree of the solid phase photochemical reaction with BPO depends on the choice of the SAM molecule.

**C. Reaction mechanism**

Fig. 4 shows possible mechanisms governing the SAM-controlled photochemical reaction with BPO. The reaction consists of several elementary processes: photo-induced dissociation of BPO molecules to generate phenyl radicals and subsequent attachment to the graphene surface. Among these processes, there are two possible pathways for the BPO dissociation: namely, direct photodissociation of BPO [Fig. 4(a)], and photocatalytic dissociation via photo-induced electron transfer from graphene to BPO [Fig. 4(b)].[30] The UV light source used in this study contains wavelengths that are absorbed by BPO (< 350 nm),[50] which indicates that a direct photodissociation pathway is likely be involved in the present case. Therefore, the controllability of the molecular gating method is verified by examining both mechanisms.

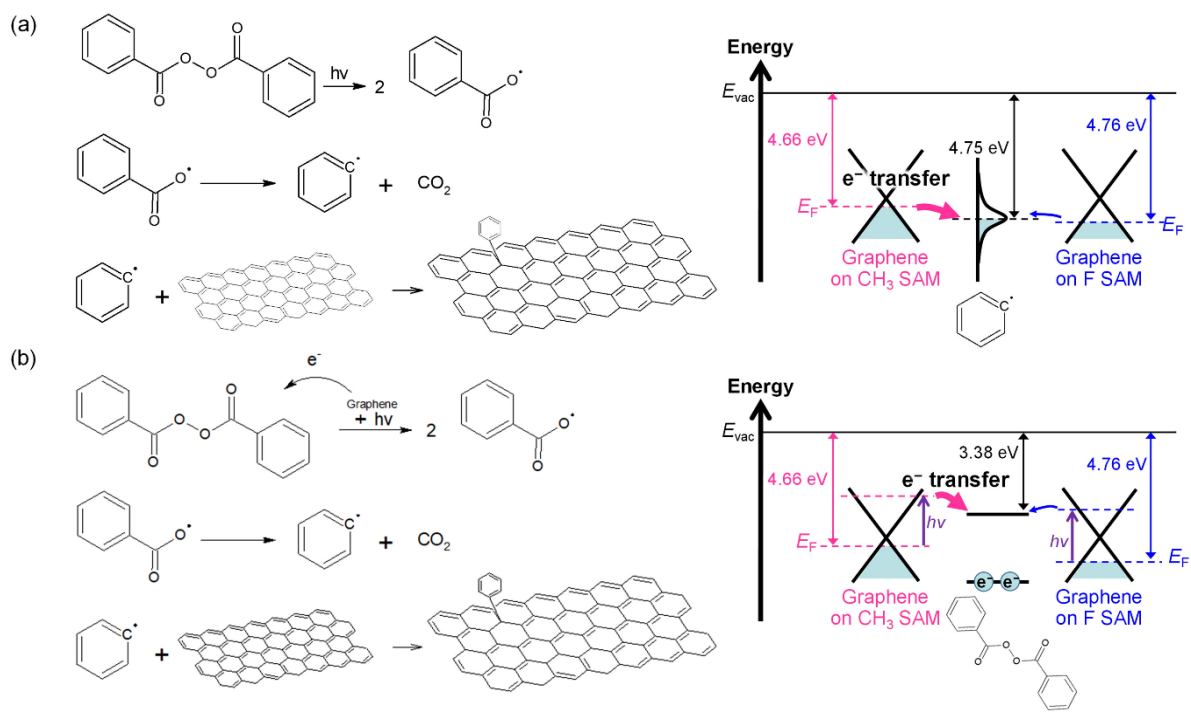

Fig. 4 Mechanisms of the SAM-controlled solid phase photochemical reaction between graphene and BPO. Possible elementary processes (left) and energy diagram (right) for (a) direct photodissociation of BPO and (b) photocatalytic dissociation of BPO. The photocatalytic pathway is less probable in the present case.

In the direct photodissociation pathway, the attachment of phenyl radicals formed by the dissociation is known to be kinetically governed by electron transfer from graphene to the phenyl radical.[12] The Fermi level of the phenyl radical is 4.75 eV below the vacuum level.[12, 51] From vector decomposition analysis, the median values of $E_F$ before the reaction were extracted as −0.10 and −0.19 eV from the Dirac point for graphene flakes formed on $CH_3$- and F-SAM, respectively, where the negative signs indicate p-type doping. The Dirac point of graphene is located 4.57 eV below the vacuum level.[52] These values can be used to construct an energy diagram, as shown in Fig. 4(a). Hence, electron transfer from graphene on F-SAM is inefficient because the Fermi level of graphene is slightly lower than that of a phenyl radical, which explains the experimental results. Thus, the controllability can be attributed to the Fermi-level-controlled efficacy of the attachment of phenyl radicals to graphene.

In the photocatalytic dissociation pathway,[30] the electron transfer from photoexcited graphene to the lowest unoccupied molecular orbital (LUMO) of adjacent BPO molecules is the rate-limiting process. The lower edge of the LUMO of BPO physisorbed on graphene was calculated to be 1.19 eV above the Dirac point,[53] and an energy diagram was constructed as Fig. 4(b). Considering the energy range of the light source used in this study (see Fig. S1 in the ESI), the photoexcited electrons in both graphene flakes on $CH_3$- and F-SAM fill energy states higher than the LUMO edge of BPO. However, we have observed that the flakes on F-SAM are rather unreactive toward solid-phase photochemical reaction with BPO. This fact indicates that the photocatalytic mechanism rarely occurred in our samples. In our experiments, the UV light was irradiated from the top of the sample, where the light passes through in the order a BPO film, graphene, a SAM, to a $SiO_2$/Si substrate. Therefore, the light should be largely absorbed in the topmost BPO layer. As a result, the portion of the light that is absorbed by the graphene layer should be very limited. Therefore, the photocatalytic mechanism rarely occurred in our experimental setup.

D. Other possible factors

The degree of chemical modification of graphene is known to be also controllable by strain,[54-56] but the present results are ascribable to the carrier doping effect as explained below. The $\Delta\omega_G$ value solely from the strain effect (i.e., the $\Delta\omega_G$ value corresponding to no doping condition) is also extractable from as-measured $\omega_G$ values as $b'\cos\vartheta_Y$. The corresponding $\omega_G$ value is obtained by $\omega_G^0 + b'\cos\vartheta_Y$. The corrected $\Delta(A_D/A_G)$ value was found to possess no clear correlation with the $\omega_G$ value solely from the strain effect (see Fig. S5 in the ESI). This fact further supports that the controllability is a result of the carrier doping effect.

Even without the BPO layer, UV irradiation can induce the Raman D band via direct photo-oxidation of graphene; however, the direct photo-oxidation was found to exert almost no effect on the results because of its low reactivity. For the direct photo-oxidation, SAM-modified substrates with graphene flakes were irradiated with the same UV irradiation system as was used for the solid phase photochemical reaction; we employed a longer irradiation period of 30 min and a higher irradiance at the sample of 2.6 W cm$^{-2}$. Because photo-oxidation in air is thermally activated and can be enhanced by humidity,[13] a thermo-hygrostat (Eyela, KCL-2000A) was used to increase the ambient temperature and humidity to 50 °C and 40 g m$^{-3}$, respectively. Photo-oxidation by the green Raman laser is known to be a very slow process,[36] thus no special care was taken for the direct photo-oxidation, unlike the solid phase photochemical reaction with BPO. The same data-correction procedure was taken for the direct photo-oxidation. However, despite the harsher reaction conditions (the three-fold irradiation period, the three-fold irradiance, etc.), the corrected Δ($A_D$/$A_G$) values were found to be still comparable to those of the solid phase photochemical reaction in the low reactivity (i.e., high hole-concentration) region in Fig. 3(d) (see Fig. S6 in the ESI). This fact indicates that the results of the solid phase photochemical reaction with BPO were almost unaffected by the direct photo-oxidation of graphene.

## 5 Conclusion

Molecular gating can be used to modify the charge carrier concentration in 2D materials by modifying the supporting substrate surfaces with polar SAMs. This method offers control over chemical reactions at the surfaces of 2D materials. Specifically, control was achieved over a solid phase photochemical reaction of graphene with BPO. The chemical reaction was controlled by choosing the molecular species with an appropriate electric dipole orientation to form the SAM. Possible mechanisms of the control are discussed by considering the elementary processes driving the reaction and can be summarized as follows. Molecular gating controls the Fermi level, which regulates electron transfer from graphene to BPO-derived phenyl radicals, and from photoexcited graphene to BPO molecules. This kinetically limits the solid phase photochemical reaction with BPO.

The molecular gating method does not require fabrication and operation of FETs and offers a simple method for controlling chemical reactions. Particularly, in the case of liquid phase reactions, the FET gating method usually requires a wire coating to prevent unwanted electrochemical reactions on the wire surfaces. This complexity in device preparation can be avoided by molecular gating because external (gate) voltages are not required. In this study, the molecular gating method was effective for controlling a representative solid-phase reaction involving an archetypal 2D material, graphene. This methodology could be used to control various chemical reactions on the surfaces of many 2D materials.

## Conflicts of interest

There are no conflicts to declare.

## Acknowledgements

This work was supported in part by the Special Coordination Funds for Promoting Science and Technology from the Ministry of Education, Culture, Sports, Science and Technology of Japan; JSPS KAKENHI Grant Numbers JP26107531, JP16H00921, JP17H01040, and JP19H02561; and JST, PRESTO Grant Number JP17939060.